\documentclass{ifacconf}
\usepackage{graphicx}      
\usepackage{natbib}
\usepackage{color}        
\usepackage{amssymb,amsmath}
\usepackage{accents}
\usepackage{enumitem}
\usepackage{todonotes}
\newtheorem{remark}{Remark}
\newtheorem{proposition}{Proposition}

\newcommand{\norm}[1]{\left\lVert#1\right\rVert}

\def\rea{\mathbb{R}}

\newcommand{\inv}[1]{{#1}^{-1}}
\newcommand{\pdev}[2]{\frac{\partial#1}{\partial#2}}
\newcommand{\ubar}[1]{\underaccent{\bar}{#1}}

\begin{document}
\begin{frontmatter}
\title{{Passivity-based control of mechanical systems with linear damping identification}}

\author{Carmen Chan-Zheng\thanksref{footnoteinfo}} 
\author{, Pablo Borja, and} 
\author{Jacquelien M.A. Scherpen}
\thanks[footnoteinfo]{The work of Carmen Chan-Zheng is sponsored by the University of Costa Rica.}
\address{Jan C. Willems Center for Systems and Control\\
	Engineering and Technology Institute Groningen (ENTEG)\\
	University of Groningen, Nijenborgh 4, 9747 AG Groningen,\\
	The Netherlands\\
	E-mail:\{c.chan.zheng, l.p.borja.rosales, j.m.a.scherpen\}@rug.nl.}

\begin{abstract}       
We propose a control approach for a class of nonlinear mechanical systems to stabilize the system under study while ensuring that the oscillations of the transient response are reduced. The approach is twofold: (i) we apply our technique for linear viscous damping identification to improve the accuracy of the selected control technique, and (ii) we implement a passivity-based controller to stabilize and reduce the oscillations by selecting the control parameters properly in accordance with the identified damping. Moreover, we provide theoretical analysis for a particular passivity-based control approach on its effectiveness for reducing such oscillations. Also, we validate the methodology by implementing it experimentally in a planar manipulator.
\end{abstract}
\end{frontmatter}
\section{Introduction}
The port-Hamiltonian (pH) framework is suitable to model a large class of systems from different physical domains, which passivity property can be verified via the derivative of the Hamiltonian (see \cite{duindam2009modeling}). This property is essential for implementing passivity-based control (PBC) techniques that guarantee stability properties for the closed-loop system via i) energy shaping and ii) damping injection (see \cite{ortega2013passivity}). However, ensuring stability properties may not be enough for practical implementation purposes as several cases are essential to ensure a prescribed performance in the transient response, e.g., reducing the oscillations. In this line of research, we find the results of \cite{wesselink2019saturated,hamada2020passivity,chan2020tuning,dirksz2013tuning,woolsey2004controlled,keppler} that provide tuning guidelines (or design procedures) for particular classes of PBC approaches to guarantee---in addition to stability---the desired performance in terms of the oscillations in the transient response of the system. 

Customarily,  models of mechanical systems neglect the natural damping of the system to simplify the analysis. Nonetheless, omitting this parameter diminishes the accuracy of the mentioned tuning guidelines as it determines the required damping to be injected. Thus, proper identification of the natural damping is essential for the implementation of these guidelines. A detailed assessment for damping identification methods based on the well-established modal analysis (MA) can be found in \cite{adhikari2001identification}. The main disadvantage of MA is that it requires different equipment types, e.g., multiple sensors and diverse equipment for acquiring and processing data. Another identification procedure is found in \cite{miranda2018experimental}, where it identifies a more extensive set of parameters---including the natural damping---without using acceleration data for a two-degree-of-freedom (DoF) planar manipulator. However, the procedure becomes complex due to avoiding acceleration data and identifying a sizeable set of parameters.  On the other hand,  \cite{liang2007damping} proposes a simpler damping identification method based on the energy of the system that requires position, velocity, and acceleration data. However, this method is restricted to diagonal and constant mass inertia matrices.

The main contribution of this paper is a control methodology that guarantees the stability of the closed-loop system while reducing the oscillations exhibited during the transient response of a class of mechanical systems where the damping characterization is a challenge. The methodology is twofold: we first propose a damping identification method that uses limited equipment. For this, we extend the approach of \cite{liang2007damping} to identify the linear damping of a larger class of mechanical systems, including those with a non-constant inertia matrix. Secondly, we implement two PBC approaches--one for fully-actuated mechanical systems and the other for underactuated mechanical systems--where we select the gains of the controllers according to the identified linear damping.  An additional contribution of this paper is the analysis of the PBC approach discussed in \cite{wesselink2019saturated} regarding its effectiveness in oscillations reduction. 

The remainder of this paper is structured as follows: in Section \ref{prel}, we describe the system under study. We present the damping identification method in Section \ref{identification_method}. In Section \ref{pipbc}, we describe the details of PI-PBC approaches with the oscillation reduction analysis. In Section \ref{expres}, we apply our methodology to an experimental case study. We finalize this manuscript with some concluding remarks in Section \ref{concl}. 

\textbf{Notation}: We denote the $n\times n$ identity matrix as $I_n$ and the $n\times m$ matrix of zeros as $0_{n\times m}$. For a given smooth function $f:\rea^n\to \rea$, we define the differential operator $\nabla_x f:=(\frac{\partial f}{\partial x})^\top$ and $\nabla^2_x f:=\frac{\partial^2 f}{\partial x^2}$. The sub-index in $\nabla$ is omitted when clear from the context. For a vector $x\in\rea^{n}$, we say that $A$ is \textit{positive definite (semi-definite)}, denoted as $A>0$  ($A\geq0$), if $A=A^{\top}$ and $x^{\top}Ax>0$ ($x^{\top}Ax\geq0$) for all $x\in \rea^{n}-\{0_{n} \}$ ($\rea^{n}$). For $A\geq0$ and $A>0$,  we define the weighted Euclidean norm as $\lVert x \rVert_{A}:=\sqrt{x^{\top}Ax}$. For $A=A^\top$, we denote by $\bar{\lambda}(A)$ and by $\ubar{\lambda}(A)$ as the maximum eigenvalue and the minimum eigenvalue of $A$, respectively. Given a distinguished element $x_\star\in\rea^n$, we define the constant matrix $B_\star:=B(x_\star)\in \rea^{n\times m}$. We denote $\rea_+$ as the set of positive real numbers and $\rea_{\geq0}$ as the  set $\rea_+ \cup \{0\}$. 

\textbf{Caveat:} The conference version of this manuscript contains a broken video link; therefore, we have updated such a link in the current version.
\section{Problem Setting}\label{prel}
Consider a mechanical system described by
\begin{equation}\label{sysmec}
	\begin{split}
		&\begin{bmatrix}
			\dot{q}\\ \dot{p}
		\end{bmatrix}= \begin{bmatrix}
			0_{n\times n}&I_n \\-I_n&-D(q,p)
		\end{bmatrix}\begin{bmatrix}
		\nabla_q H(q,p)\\\nabla_p H(q,p)
	\end{bmatrix}+ \begin{bmatrix}
			0_{n\times m}\\G
		\end{bmatrix}u,\\
		&H(q,p)=\frac{1}{2}p^\top \inv{M}(q)p + V(q), \quad y= G^{\top}\dot{q},
	\end{split}
\end{equation} 
where $q,p \in \mathbb{R}^n$ are the generalized positions and momenta vectors, respectively; $u,y\in \mathbb{R}^m$ are the control and passive output vector, respectively; $m\leq n$; ${D: \mathbb{R}^n\times\mathbb{R}^n \to \mathbb{R}^{n \times n}}$ is the positive definite natural damping matrix; $H:\mathbb{R}^n\times\mathbb{R}^n \to \mathbb{R}_{\geq 0}$ is the Hamiltonian of the system; $M:\mathbb{R}^n \to \mathbb{R}^{n \times n}$ is the positive definite mass-inertia matrix; $V:\mathbb{R}^n\to \mathbb{R}_{\geq 0}$ is the potential energy; and $G\in \mathbb{R}^{n\times m}$ is the constant input matrix defined as 
\begin{equation*}
	G:=\begin{bmatrix}
		0_{r\times m}\\G_1
	\end{bmatrix}
\end{equation*}
where  $G_1\in \rea^{m\times m}$ is a full rank constant matrix, and $r:=n-m$.

Let $q:=[q_u^\top,q_a^\top]^\top$, where $q_u\in\rea^{r}$ and $q_a\in\rea^m$ correspond to the unactuated and the actuated coordinates, respectively.  Then, we identify the set of assignable equilibria for \eqref{sysmec} given by
$$
	\mathcal{E} = \left\lbrace (q,p)\in\rea^{n}\times\rea^{n} \mid \nabla_{q_u} V(q) = 0_{r}, p=0_{n}\right\rbrace.
$$

We proceed to formulate the problem under study: propose a control methodology such that in closed-loop with \eqref{sysmec} the oscillations in the transient response are reduced.

The control methodology is twofold: (i) we propose an identification method to determine the linear damping of the system in an easy manner, and (ii) we implement two particular PI-PBC approaches, where we select the gains according to the identified damping in (i). In the sequel, we proceed to describe each step in detail.

\section{Energy-based damping identification method}\label{identification_method}
In this section, we propose an energy-based damping identification (EBDI) method to estimate the linear damping for a class of mechanical systems similar to the approach in \cite{liang2007damping}. However, the latter reference derives the identification algorithm from the Euler-Lagrange framework and considers only systems with constant and diagonal mass-inertia matrices.  In this manuscript, we extend the mentioned approach to a larger class of systems by deriving the algorithm from a pH framework. 

Additionally, the EBDI methodology considers the linear component of $D(q,p)$---i.e., the linear viscous damping---as the tuning of the PBCs described in the sequel only requires the characterization of this linear term. Moreover, without loss of generality, we consider the linear damping of the system as a constant diagonal matrix, i.e., 
\begin{equation*}
	D:=diag\{d_1,d_2,\hdots,d_n\}
\end{equation*}
with $d_1,d_2,\hdots,d_n \in \rea_+$.\footnote{Considering positive values for the diagonal entries of $D$ is not restrictive from a physical standpoint as the damping is inherent to the nature of the mechanical systems.}

Subsequently, the following proposition establishes one of the main results of this manuscript.
\begin{proposition}[EBDI methodology]
	Consider the dynamics of \eqref{sysmec}. Let
	\begin{equation}\label{reg1}
		\def\arraystretch{1.4}
		\begin{array}{rcl}
			\gamma&:=&\begin{bmatrix}
				d_1&\hdots&d_n
			\end{bmatrix}^\top,\\
			\Phi(q,\dot{q},\ddot{q})&:=&\begin{bmatrix}
				\Phi_1(q,\dot{q},\ddot{q})&\hdots& \Phi_\ell(q,\dot{q},\ddot{q})
			\end{bmatrix}^\top,\\
		   \Psi(\dot{q})&:=&\begin{bmatrix}
				\Psi_1(\dot{q})&\hdots&\Psi_\ell(\dot{q})
			\end{bmatrix}^\top,
		\end{array}
	\end{equation}
	where $\Psi(\dot{q})$ and $\Phi(q,\dot{q},\ddot{q})$ are constructed with $\ell\in\mathbb{N}$ sets of measurements, and
	\begin{equation}\label{reg2}
		\def\arraystretch{1.6}
		\begin{array}{rcl}
			\Phi_h(q,\dot{q},\ddot{q})&:=&\begin{bmatrix}
				\phi_1(q,\dot{q}_1,\ddot{q})&\hdots&\phi_n(q,\dot{q},\ddot{q})
			\end{bmatrix}_h^\top,\\
			\Psi_h(\dot{q})&:=&-diag\left\{	\int\dot{q}_1^2~dt,\hdots,	\int\dot{q}_n^2~dt\right\}_h, \\
		\end{array}
	\end{equation}
	with $h=1,\hdots,\ell$, and
	\begin{equation}\label{phi}
		\begin{array}{rl}
			\phi_k(q,\dot{q},\ddot{q}):=&\int \Big[\dot{q}_{k} e_k^\top \dot{M}(q)\dot{q}+\dot{q}_{k}e_k^\top M(q)\ddot{q}+\dot{q}_{k}\nabla_{q_k}V(q)\\
			&-\frac{1}{2}\dot{q}_{k}\dot{q}^\top\pdev{M(q)}{q_k}\dot{q}-\dot{q}_k e_k^\top Gu\Big]~dt,
		\end{array}
	\end{equation}
	where $k=1,\hdots,n$; and $e_k$ is the $k^{th}$ element of the canonical basis of $\rea^n$.	
	Then, the optimal solution for $\gamma$, in least-square sense, becomes
	\begin{equation}\label{solution}
		\gamma=\inv{[\Psi^\top(\dot{q}) \Psi(\dot{q})]}\Psi^\top(\dot{q}) \Phi(q,\dot{q},\ddot{q}).
	\end{equation}
\end{proposition}

\begin{pf}
	From \eqref{sysmec}, $\dot{p}$ is given by
	\begin{equation}\label{pdot1}
		\begin{split}
			\dot{p}&=-\nabla_qH(q,p)-D\nabla_pH(q,p)+Gu\\
			&=-\nabla_qV(q)+\frac{1}{2}\sum^n_{i=1}e_i\dot{q}^\top\pdev{M(q)}{q_i}\dot{q}-D\dot{q}+Gu.
		\end{split}
	\end{equation}
	On the other hand, note that $\dot{p}$ can also be computed as:
	\begin{equation}\label{pdot2}
		\begin{split}
			p=M(q)\dot{q}\implies\dot{p}&=\frac{d}{dt}(M(q)\dot{q})\\
			&=\dot{M}(q)\dot{q}+M(q)\ddot{q}.
		\end{split}
	\end{equation}
	Therefore, by comparing and rearranging \eqref{pdot1} and \eqref{pdot2}, we have the following expression:
	\begin{equation}\label{pdot3}
		\begin{array}{lll}
			-D\dot{q}=\dot{M}(q)\dot{q}&+M(q)\ddot{q}+\nabla_qV(q)&\\
			&-\frac{1}{2}\sum^n_{i=1}e_i\dot{q}^\top\pdev{M(q)}{q_i}\dot{q}-Gu.
		\end{array}
	\end{equation}
	Then, \eqref{pdot3} can be decoupled into $n$-equations of the form
	\begin{equation}\label{pdot4}
				\def\arraystretch{1.6}
		\begin{array}{rl}
			-d_k\dot{q}_k=e_k^\top \dot{M}(q)\dot{q}+e_k^\top M(q)\ddot{q}+&\nabla_{q_k}V(q)\\
			-\frac{1}{2}\dot{q}^\top\pdev{M(q)}{q_k}\dot{q}-e_k^\top Gu&
		\end{array}
	\end{equation}
	with $k=1,2,\hdots,n$.
	
	By multiplying with $\dot{q}_k$ and integrating each side of \eqref{pdot4}, we get the energy expression
	\begin{equation}\label{energy}
				\def\arraystretch{1}
		\begin{array}{lll}
			-\int d_k\dot{q}^2_k~dt=	\phi_k(q,\dot{q},\ddot{q}),
		\end{array}
	\end{equation}
	where $\phi_k(q,\dot{q},\ddot{q})$ is defined as in \eqref{phi}.
	Then, by regrouping the $n$-equations \eqref{energy}, we get the following matrix form
	\begin{equation}\label{regresor}
		\Psi(\dot{q})\gamma = \Phi(q,\dot{q},\ddot{q}),
	\end{equation}
	where $\gamma$, $\Psi(\dot{q})$, and $\Phi(q,\dot{q},\ddot{q})$ are defined as in \eqref{reg1}-\eqref{reg2}.
	
	Hence, by multiplying \eqref{regresor} with the pseudo-inverse of $\Psi(\dot{q})$, we get \eqref{solution}.
	\flushright$\blacksquare$
\end{pf}
\begin{remark}
	The identification methodology explained in the current section is not limited to damping identification. Note that a mechanical system is characterized by the mass-inertia, damping, and stiffness matrices.  Thus, by following a similar approach, any of these matrices can be identified if the other two matrices are known.
\end{remark}
\begin{remark}
	The EBDI approach can be applied to either open or closed-loop systems. In general, most of the identification processes are performed in an open loop. On the other hand, a closed-loop identification is used when the open-loop is unstable or when the implementation in an open loop is not practical due to safety, production losses, among other concerns.
\end{remark}

\section{PI-PBC approaches}\label{pipbc}
In this section, we describe two PI-PBC approaches that stabilize \eqref{sysmec} at the desired equilibrium $x_\star:=(q_\star,0_n)$ with $q_\star\in\rea^n$ as the desired configuration. 

The first control law corresponds to the PI-PBC described in \cite{borja2020new,zhang2017pid} where in combination with the tuning methodology of \cite{chan2020tuning} shows a suitable approach to reduce oscillations for a class of mechanical systems. The second control law is found in \cite{wesselink2019saturated} where it describes a modified PI-PBC methodology that effectively reduces the oscillations in the transient response of some coordinates by injecting an additional damping term related to the unactuated coordinates. However, such an effect is demonstrated only via experimental results. Hence, in this paper, we also provide an analytical discussion about such behavior. 

Additionally, the mentioned control laws render the target dynamics into the form 
\begin{equation}\label{cldyn}
	\begin{bmatrix}
		\dot{q}\\ \dot{p}
	\end{bmatrix}=F(q,p)\nabla H_d(q,p) 
\end{equation} 
where $F:\rea^n\times \rea^n\to\rea^{2n\times 2n}$ and $H_d:\rea^n\times\rea^n\to\rea$ are defined accordingly to each methodology discussed in the sequel.

\subsection{The PI-PBC approach}
The PI-PBC approach from \cite{borja2020new,zhang2017pid} does not require the solution of partial differential equations and its control parameters admit a physical interpretation. We summarize such a result in Proposition \ref{prop1}.
\begin{proposition}[PI-PBC]\label{prop1}
	Consider the pH system \eqref{sysmec}. Then, define the control law
	\begin{equation}\label{pid}
		u=-K_P \dot{q}_a-K_I(q_a- q_{a\star}-\inv{K_I}\nabla_{q_a} V(q_{a\star})),
	\end{equation}
	where the gains $K_P$ and $K_I \in \mathbb{R}^{m\times m}$ satisfy $K_I>0$, $K_P\geq0$; and $q_{a\star} \in\rea^m$ corresponds to the desired configuration for the actuated coordinates.
	Then, the closed-loop system has a stable equilibrium at $x_\star$ if there exists $K_I>0$ such that
	\begin{equation*}
		\nabla^2 V(q_\star)+GK_IG^\top>0.
	\end{equation*}
	Furthermore, the closed-loop system \eqref{sysmec}, \eqref{pid} has the form \eqref{cldyn}, with 
	\begin{equation*}
		\begin{array}{rcl}
		H_d(q,p)&:=&H(q,p)+\frac{1}{2}\lVert q_a- q_{a\star}-\inv{K_I}\nabla_{q_a} V(q_{a\star})\rVert^2_{K_I}\\
		F(q,p)&:=&\begin{bmatrix}0_{n\times n}&I_n\\-I_n & -D(q,p)-GK_PG^\top\end{bmatrix}.
				\end{array}
	\end{equation*}
$\square$
\end{proposition} 
\begin{remark}
The control approach described in \cite{borja2020new,zhang2017pid} contains a derivative action. However, we omit its use in this manuscript as it is not required to reduce the oscillations \eqref{sysmec}. Moreover, the addition of this action requires extra damping injection---i.e., a larger $K_P$ gain---resulting in a slower convergence (see \cite{chan2020exponential}). Thus, we assume that the PI-PBC is suitable to guarantee the stability of the desired equilibrium of the closed-loop system.
\end{remark}

\subsubsection{A tuning guideline for the PI-PBC}
\hfill
\vspace{1mm}

The authors in \cite{chan2020exponential} show that the PI-PBC in Proposition \ref{prop1} preserves the mechanical structure---additionally to the pH structure---in the closed-loop system, which is essential to establish the tuning guidelines presented in \cite{chan2020tuning}
The latter reference proposes a procedure to choose the gains $K_P$ and $K_I$ such that the transient response of the closed-loop system contains minimum oscillations\footnote{A similar procedure for choosing the gains of the PI-PBC approach can be found in \cite{hamada2020passivity}. However, such a procedure is applicable for closed-loop systems with dynamic extension, which is not the case of the closed-loop system \eqref{sysmec}, \eqref{pid}.}. We summarize this result in Proposition \ref{prop2}.
\begin{proposition}\label{prop2}
	Consider the pH system \eqref{sysmec} in closed-loop with \eqref{pid}. If
	\begin{equation}\label{tuningrule}
		\begin{array}{rl}
			\ubar{\lambda}(GK_PG^\top+D_\star)^2\geq
			4\bar{\lambda}(GK_IG^\top+\nabla^2V_\star)&\bar{\lambda}(M_\star)
		\end{array}
	\end{equation}
	holds, then, the closed-loop system \eqref{sysmec}, \eqref{pid} does not exhibit oscillations in its transient response.
\end{proposition}
	$\square$

\subsection{The modified PI-PBC approach}\label{modiPI}
For fully-actuated systems, the PI-PBC \eqref{pid} can modify the potential energy and the damping of \textit{every coordinate}. Thus, the condition \eqref{tuningrule} can be satisfied straightforwardly. However,  for underactuated systems, the performance may be constrained by the unactuated coordinates. Hence, condition \eqref{tuningrule} may not be applicable when the linear damping of the unactuated coordinates is too small. To overcome this obstacle, instead of the PI-PBC \eqref{pid}, the authors in \cite{wesselink2019saturated} propose the modified PI-PBC, which we summarize in Proposition \ref{prop3}.
\begin{proposition}[Modified PI-PBC]\label{prop3}
	Consider the pH system \eqref{sysmec} with 
	$$D:=diag\{D_u , D_a\},$$ 
	where $D_u\in\rea^{r\times r}$ and $D_a\in\rea^{m\times m}$ are constant positive definite matrices representing the natural damping of the unactuated and the actuated coordinates, respectively; and define the control law
	\begin{equation}\label{pimod}
		u=-K_IG_1^\top(q_a-q_{a\star})-K_{Pa}G_1^\top\dot{q}_a-K_{Pu}\dot{q}_u
	\end{equation}
	where $K_I\in\rea^{m\times m}$ is a positive definite matrix that ensures that the potential energy of the closed-loop system has an \textit{isolated minimum} at $x_\star$; $K_{Pa} \in \rea^{m\times m}$ is positive definite; and $K_{Pu}\in\rea^{m\times r}$. These matrices satisfy	\begin{equation}\label{condpimod}
		\begin{array}{lll}
			 D_a+G_1K_{Pa}G_1^\top-\frac{1}{4}G_1K_{Pu} \inv{D_u}K_{Pu}^\top G_1^\top>0.
		\end{array}
	\end{equation}

	Then, the closed-loop system has a locally asymptotically stable equilibrium at $x_\star$. Furthermore, the closed-loop system \eqref{sysmec},\eqref{pimod} has the form \eqref{cldyn} with 
	\begin{equation}\label{clmatrices2}
		\def\arraystretch{1}
		\begin{array}{rl}
			H_d(q,p)&:=H(q,p)+\frac{1}{2}\norm{q_a-q_{a\star}}^2_{K_I},\\
			F(q,p)&:=\begin{bmatrix}0_{n\times n}&I_n\\-I_n & J_2-R_2\end{bmatrix},\\
			R_2&:=\begin{bmatrix}D_u&\frac{1}{2}(G_1K_{Pu})^\top\\\frac{1}{2}G_1K_{Pu} & D_a+G_1K_{Pa}G_1^\top\end{bmatrix},\\
			J_2&:=\begin{bmatrix}0_{r\times r}&\frac{1}{2}(G_1K_{Pu})^\top\\-\frac{1}{2}G_1K_{Pu} & 0_{m\times m}\end{bmatrix}.
		\end{array}
	\end{equation}
	$\square$
\end{proposition}

\begin{remark}\label{tuningrule2}
	Note that \eqref{condpimod} holds if 
	\begin{equation}\label{t2}
		\begin{array}{rl}
			4\ubar{\lambda}(D_a+G_1K_{Pa}G_1^\top)>
			\bar{\lambda}(G_1K_{Pu} \inv{D_u}K_{Pu}^\top G_1^\top),
		\end{array}
	\end{equation}
	which eases the verification during the practical implementation.
\end{remark}
\begin{remark}\label{r5}
	The pH and mechanical structure are preserved even with the addition of the extra term $K_{Pu}\dot{q}_u$, which modifies the interconnection structure of \eqref{sysmec}. Furthermore, with $K_{Pu}=0_{m\times r}$, we recover the original PI-PBC of \eqref{pid}.
\end{remark}

\begin{remark}
	The tuning guideline in Proposition \ref{prop2} is no longer applicable to the modified PI-PBC since the block $(2,2)$ of $F(q,p)$ in \eqref{clmatrices2} is not symmetric. 
\end{remark}
\begin{remark}
	Note that the conditions \eqref{tuningrule} and \eqref{t2} require knowledge on the natural damping of the system. Therefore, the accuracy of these conditions improves with a proper identification methodology.
\end{remark}

\vspace{1mm}

\subsubsection{An analysis for the modified PI-PBC}
\hfill
\vspace{1mm}

The authors of \cite{wesselink2019saturated} have shown via experimental results that the controller \eqref{pimod} increases the rate of convergence, which may minimize the oscillations in the transient response of some coordinates. However, no theoretical background is provided that explains such  behavior. Thus, in this paper, we extend this result by providing an analysis of the mentioned effect. 

The effect of the $K_{Pu}$ gain on the rate of convergence can be explained via the results of \cite{chan2020exponential}. In this reference, the authors demonstrate that the trajectories of the closed-loop system \eqref{sysmec},\eqref{pimod} converge exponentially and provide an expression for the rate of convergence in terms of the gains of the controller, which we summarize in the following proposition.
\begin{proposition}\label{prop4}
	Consider the closed-loop system \eqref{sysmec}-\eqref{pimod}. Then,
	\begin{itemize}
		\item $x_\star$ is an exponentially stable equilibrium point with Lyapunov function
		\begin{equation*}
			S(q,p)=H_d(q,p)+ \epsilon p^\top A^\top(q)\nabla_q V_d(q),
		\end{equation*}
		where $\epsilon\in\rea_+$, $V_d(q):=V(q)+\frac{1}{2}\norm{q_a-q_\star}^2_{K_I}$, and $A(q)\in\rea^{n\times n}$ is the upper Cholesky factor of  $
		\inv{M}(q)=A^\top(q)A(q).
		$
		\item The trajectories of the closed-loop system converge to the desired equilibriun $x_\star$ with a rate of convergence given by
		\begin{equation}\label{convergence}
			\frac{\beta_{\max}\mu}{1+\epsilon\norm{A(q)}\beta_{\max}},
		\end{equation}
		where $\beta_{\max},\mu\in\rea_+$.
	\end{itemize} 
\end{proposition} 
$\square$

It can be seen that the rate of convergence  \eqref{convergence} can be increased by augmenting $\mu$. This term is defined as the minimum eigenvalue of the matrix\footnote{We omit the arguments for brevity, see the proof in \cite{chan2020exponential} for further details.}
\begin{equation}\label{upsilonsym2}
	\arraycolsep=0.8pt \def\arraystretch{1.4}
	\begin{array}{lll}
		\Upsilon_{sym}&= \begin{bmatrix}
			\epsilon\inv{M}&\frac{\epsilon}{2}[A\mathcal{F}-\dot{A}]\\
			\frac{\epsilon}{2}[\dot{A}^\top-A^\top \mathcal{F}^\top]&\mathcal{D}-\epsilon(A^\top \nabla_q^2V_dA)
		\end{bmatrix}
	\end{array},
\end{equation} 
where $\mathcal{D}:=A^\top (q) R_2 A(q)$, $\mathcal{F}: = A^\top (q) (R_2-J_2) A(q)-J_3(q,p)$, and $R_2,J_2$ are defined in \eqref{clmatrices2}\footnote{We omit the definition of $J_3(q,p)$ as this term is irrelevant for the current analysis.}.

Note that $\mu$ is related to $K_{Pu}$ via the matrix $\mathcal{F}$. The effect of $K_{Pu}$ on $\mu$ can be explained by the Gershgorin circle theorem, where it provides a tool to estimate the eigenvalues of a matrix. We summarize this result in Proposition \ref{prop5} (see \cite{horn2012matrix} for further details).
\begin{proposition}\label{prop5}
	Let $\upsilon_{ij}$ the $ij$-element of $\Upsilon_{sym}$ with $i,j=1,\hdots,2n$. Moreover, consider the $2n$ Gersgorin discs
	\begin{equation*}
	\left\lbrace z\in \mathbb{C}:|z-\upsilon_{ii}|\leq r_i\right\rbrace.
	\end{equation*}
	where $r_i:=\sum^{2n}_{j=1,j\neq i}|\upsilon_{ij}|$ is the radius of the disc $i$, with $i=1,\hdots,2n$.	
	Then, the eigenvalues of $\Upsilon_{sym}$ are in the union of the Gersgorin discs.
\end{proposition}
$\square$

With the additional damping term $K_{Pu}$, the radius of some of the Gersgorin discs increases. Consequently, the location of the eigenvalues of $\Upsilon_{sym}$ may vary. Moreover, this yields that some eigenvalues--including $\mu$--may be farther from the imaginary-axis; and thus,  yielding a faster rate of convergence.

Hence, by increasing the values of $K_{Pu}$, we may obtain a faster rate of convergence; consequently,  the transient response of the closed-loop system may contain fewer oscillations for some coordinates.

\section{Case study: a 2-DoF planar manipulator}\label{expres}
\begin{figure}[t]
	\centering
	\includegraphics[width=0.5\columnwidth]{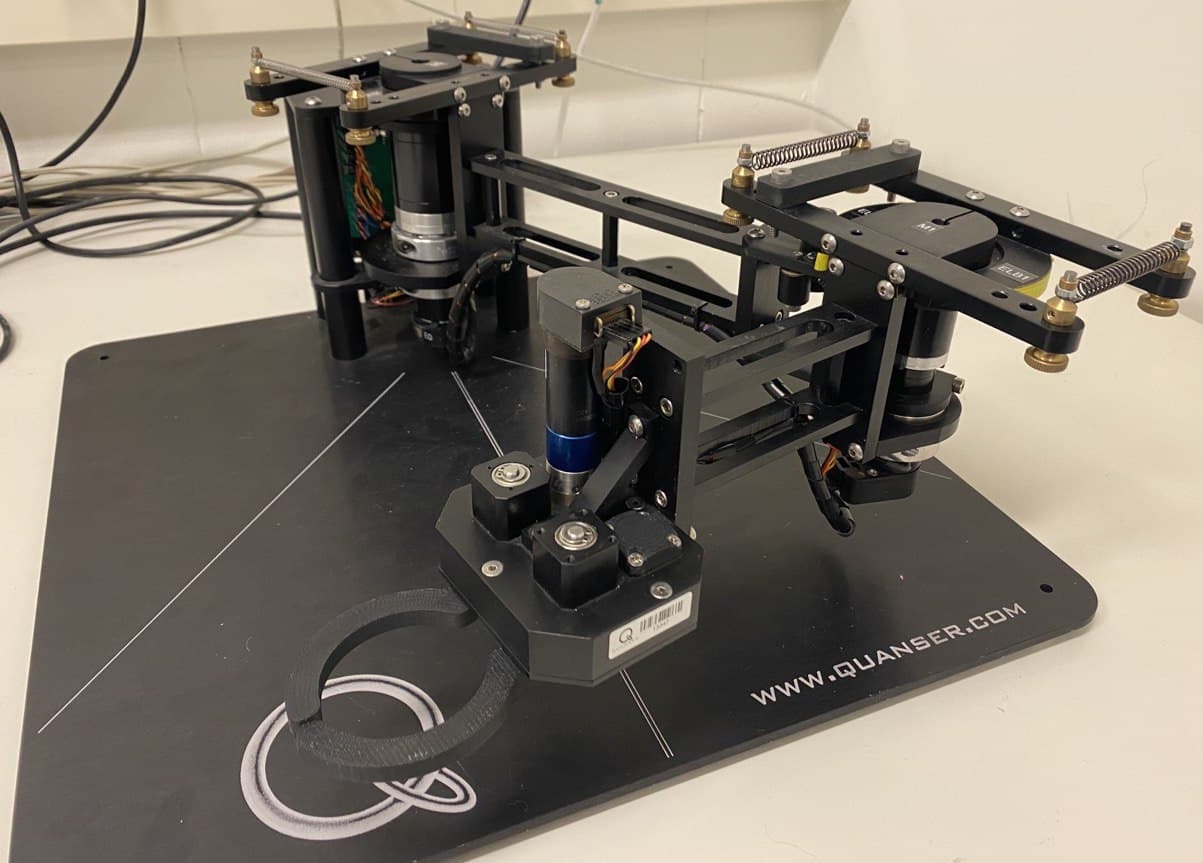}
	\caption{Quanser 2-DoF Planar Manipulator}\label{man}
\end{figure}

This section illustrates our approach for identifying the linear damping of the system and reducing oscillations in the closed-loop system.  Towards this end, we consider a 2-DoF planar manipulator as a case of study, which can be found in Fig.~\ref{man}. This experimental setup can be configured with (i) rigid joints (fully-actuated case) and (ii) flexible joints (underactuated case).  For further details, see \cite{quanser} for the reference manual.

Our experiments are performed in two steps:
\begin{enumerate}
	\item  We employ the EBDI methodology described in Section \ref{identification_method} to identify the linear damping of the system. Towards this end, we perform a closed-loop system identification by considering the PI-PBC approach of Proposition \ref{prop1} for the controller $u$. The gains are chosen according to the configuration:
	\begin{itemize}
		\item Rigid joints:  $q_{\star}=[0.6,0.8]^\top$, $K_I=diag\{30,10\}$, and $K_P=0_{2\times 2}$.
		\item Flexible joints:  $q_{\star}=[0.6,0.8,0.6,0.8]^\top$, $K_I=diag\{30,10\}$, and $K_P=diag\{0,0.5\}$.
	\end{itemize}
	Moreover, the matrices in \eqref{regresor} are constructed with the measurements of five tests ($\ell=5$).
	\item  We select a PI-PBC approach---\eqref{pid} for rigid configuration and  \eqref{pimod} for flexible configuration---to control and reduce the oscillations by selecting the gains properly in accordance with the identified linear damping. 
\end{enumerate}

\subsection{Experimental results of the rigid joints configuration}
The mathematical model of the fully-actuated configuration is described as in \eqref{sysmec} with $n=2$, $m=2$, the position and momenta of each link are denoted with $q_i$ and $p_i$ ($i=1,2$), respectively. Moreover, $q_a=[q_1,q_2]^\top$, $V(q)=0$, $G_1=diag\{1,1.6667\}$, and $M(q)=M_l(q_2)$ with
\begin{equation*}
	\def\arraystretch{1.5}
	\begin{array}{rl}
		M_l(q_2)&:=\begin{bmatrix}
			a_1+a_2+2b\cos(q_2)&a_2+b\cos(q_2)\\
			a_2+b\cos(q_2)&a_2
		\end{bmatrix},
	\end{array}
\end{equation*} 
where $a_1=0.1547,~a_2=0.0111$, $b=0.0168$.

With $D=diag\{d_1,d_2\}$, then, the values of $d_1$ and $d_2$ are obtained through the solution of \eqref{solution} which corresponds to
\begin{equation*}
	\gamma=[1.5964~0.6971]^\top.
\end{equation*}

Then, we employ the PI-PBC \eqref{pid} with different cases of gains selection as shown in Table \ref{exp1}.
\begin{table}[t]
	\centering
	\caption{Gains for the PI-PBC} \label{exp1}
	\begin{tabular}{cccc}
		\hline
		& Case 1      & Case 2                 & Case 3            \\ \hline
		$K_P$ & diag\{0.1,0.1\} & diag\{3.2045,1.4774\} & diag\{3,1.4774\} \\
		$K_I$ & diag\{30,10\} & diag\{30,10\} & diag\{30,10\} \\
		\hline
	\end{tabular}
\end{table}
\begin{figure}[t]
	\centering
	\includegraphics[width=0.9\columnwidth]{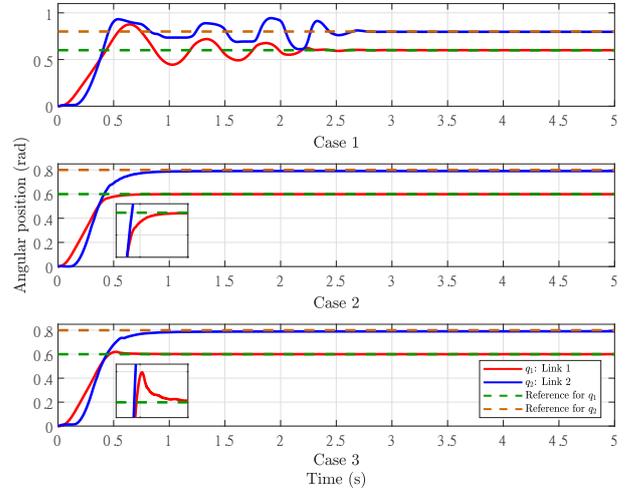}
	\caption{Trajectories of the angular positions for different cases.}\label{rigid_data}
\end{figure}
To show the effectiveness of the tuning methodology of Proposition \ref{prop2}, we first obtain a response with the gains as shown in Case 1. These gains are selected without any tuning guideline for comparison purposes, and note that the response is highly oscillatory in Case 1 of Fig~\ref{rigid_data}.  Then, we apply the tuning rule \eqref{tuningrule} with $D_\star=diag\{\gamma\}$ and fixing $K_I$. Thus, the calculated $K_P$ is shown in Table \ref{exp1} and the response is recorded in Case 2 of Fig.~\ref{rigid_data}.  Note that the number of oscillations is reduced substantially in comparison with Case 1. Furthermore, to show the accuracy of the tuning rule \eqref{tuningrule}, note that by varying slightly $K_P$ as seen in Case 3, we obtain an underdamped response. A video of the experiments can be found in\\
\texttt{https://www.youtube.com/watch?v=HKDoFQcu3mA}.

\subsection{Experimental results of the flexible joints configuration}
The manipulator with flexible joints is described as in \eqref{sysmec} with $n=4$, $m=2$; the vector $q_{1}$ (resp. $q_2$) corresponds to the position of the link 1 (resp. link 2); the vector $q_{3}$ (resp. $q_4$) corresponds to the position of the motor of link 1 (resp. motor of link 2); the vector $p_{1}$ (resp. $p_2$) corresponds to the momenta of the link 1 (resp. link 2); and the vector $p_{3}$ (resp. $p_4$) corresponds to the momenta of the motor of link 1 (resp. motor of link 2). Moreover, let $q_a=[q_1,q_2]^\top$ and $q_u=[q_3,q_4]^\top$, we have that 
\begin{equation*}
	\def\arraystretch{1.5}
	\begin{array}{rl}
		G_1&=diag\{1,1.6667\},~ 
		V(q)=\frac{1}{2}\norm{q_a-q_u}_{K_s}^2,\\
			K_s&=diag\{8.43, 16.86\},\\
		M(q)&=\begin{bmatrix}M_l(q_2)& 0_{2\times 2}\\0_{2\times 2}& M_m\end{bmatrix},~
		M_m=diag\{0.0628, 0.0026\}.
		\end{array}
\end{equation*} 

With  $D=diag\{d_1,d_2,d_3,d_4\}$, then, the identified values correspond to:
\begin{equation*}
	\gamma=[d_1~d_2~d_3~d_4]^\top=[0.0331~0.0077~2.9758~2.8064]^\top.
\end{equation*}

Then, we employ the modified PI-PBC control \eqref{pimod} with the different cases as shown in Table \ref{expund}. 
\begin{table}[t]
	\centering
	\caption{Gains for the modified PI-PBC}\label{expund}
	\begin{tabular}{cccc}
		\hline
		& $K_{Pa}$          & $K_{Pu}$             & $K_{I}$              \\ \hline
		Case 4         & diag\{5,2\} & diag\{0,0\}   & diag\{30,10\} \\
		Case 5 & diag\{5,2\}  & diag\{1,0.01\} & diag\{30,10\}  \\ \hline
	\end{tabular}
\end{table} 

\begin{figure}[t]
	\centering
	\includegraphics[width=0.9\columnwidth]{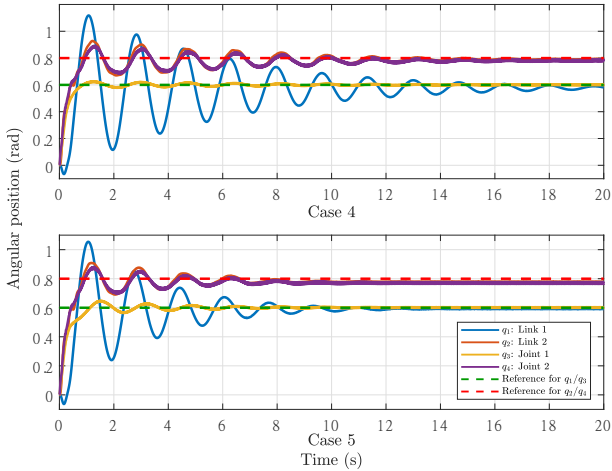}
	\caption{Trajectories of the angular positions for different cases. Top: The PI-PBC case. Bottom: The modified PI-PBC case.}\label{und_data}
\end{figure}

For comparison purposes, we first implement \eqref{pimod} without the additional damping term as seen in Case 4, which corresponds to a regular PI-PBC (see Remark \ref{r5}); the response is recorded in the top image of Fig~\ref{und_data}.  Subsequently, we apply \eqref{t2} to select adequately the $K_{Pu}$ gain as shown in Case 5; the response is given in the bottom image of Fig~\ref{und_data}. Note that---as explained in Section \ref{modiPI}---the number of oscillations for most of the coordinates reduces. A video of the flexible joint configuration experiments can be found in \texttt{https://www.youtube.com/watch?v=L7474CfhB5w}.

\section{Concluding Remarks}\label{concl}
We have demonstrated that the EBDI approach is a suitable and accurate methodology to estimate the linear damping for a class of mechanical systems; moreover, this methodology is endowed with physical interpretation since it this formulated from an energy perspective. Additionally, we have provided an analysis of the modified PI-PBC about its effectiveness in reducing oscillations for a class of nonlinear mechanical systems.
Finally, we have demonstrated via experimental results that our control methodology--i.e., the EBDI methodology in combination with the tuning guidelines of the PI-PBC or with the modified PI-PBC-- reduces the oscillations substantially in the transient response of the closed-loop system. 

\bibliography{ifacconf}             

\end{document}